\documentclass[reprint,amsmath,amssymb,aps,superscriptaddress]{revtex4-2}
\usepackage{amsmath,amssymb}
\usepackage{braket}
\usepackage{color}
\usepackage{graphicx}
\usepackage{dcolumn}
\usepackage{bm}

\usepackage{algorithmic}
\usepackage{algorithm}

\begin{document}


\title{Experimental implementation of quantum greedy optimization on quantum computer}


\author{Tadayoshi Matsumori}
\email{matsumori@mosk.tytlabs.co.jp}
\affiliation{Toyota Central R\&D Labs. Inc., Nagakute, Aichi 480-1192, Japan}
\affiliation{DENSO CORPORATION, Minato-ku, Tokyo 108-0075, Japan}
\author{Tadashi Kadowaki}
\affiliation{DENSO CORPORATION, Minato-ku, Tokyo 108-0075, Japan}
\affiliation{National Institute of Advanced Industrial Science and Technology, Tsukuba, Ibaraki 305-8568, Japan.}


\date{\today}

\begin{abstract}
	
	This paper implements a quantum greedy optimization algorithm based on the discretization of time evolution (d-QGO).
	Quantum greedy optimization, which was originally developed for reducing processing time via counterdiabatic driving, sequentially selects a parameter in the counterdiabatic term from the sensitivity analysis of energy and then determines the parameter value.
	For implementing d-QGO on a quantum computer, 
	the sensitivity analysis may become a bottleneck to find the ground state in a short time due to device and shot noise.
	In this paper, we present an improved sensitivity analysis for d-QGO that employs a sufficiently large differential interval.
	We demonstrate that d-QGO reduces the number of shots required to determine the sensitivity while maintaining the success probability.
	
\end{abstract}


\maketitle

\section{Introduction}

There is growing interest in optimization algorithms that leverage quantum dynamics and/or quantum devices. 
Several promising algorithms have emerged, including quantum annealing (QA) \cite{kadowaki1998}, adiabatic quantum computing (AQC) \cite{farhi2001}, the quantum approximate optimization algorithm (QAOA) \cite{farhi2014}, the variational quantum algorithm \cite{peruzzo2014}, and Grover adaptive search \cite{baritompa2005,gilliam2021}.
Among these algorithms, QA stands out as metaheuristics for combinatorial optimization based on quantum physics; it finds the optimum solution corresponding to the ground state of a particular Hamiltonian, such as the Ising model.
The annealing process in QA is crucial to success. 
When a system has a first-order quantum phase transition, the energy gap between the ground state and the first excited state in the Hamiltonian closes exponentially as a function of system size \cite{somma2007,jorg2010,young2010,jorg2010first}.
To find the ground state, QA thus requires a sufficiently long annealing time, which may lead to an exponential computational time due to the complexity of the problem Hamiltonian.

Counterdiabatic (CD) driving is a promising approach for accelerating the annealing process \cite{sels2017,nakahara2022}.
The concept behind CD driving is to control the fast time evolution by modifying a quantum system's Hamiltonian, which enables shortcuts to adiabaticity \cite{torrontegui2013,del2013,guery2019}.
The Hamiltonian of QA with CD driving consists of a CD term in addition to the conventional QA Hamiltonian (i.e., the transverse field Hamiltonian and the problem Hamiltonian) \cite{sels2017,takahashi2017,kolodrubetz2017,prielinger2021}.
CD driving also works with state-of-the-art hybrid quantum-classical variational algorithms, such as AQC \cite{hegade2021,hegade2022}, the QAOA \cite{yao2021,chandarana2022,wurtz2022}, and the variational quantum eigensolver \cite{zhan2021,sun2022}. 
In a previous study \cite{hegade2021}, AQC with CD driving was implemented on a quantum device and achieved high success probability (SP) in a short annealing time compared with that obtained without the CD term.

Quantum greedy optimization (QGO) \cite{kadowaki2021} is a hybrid quantum-classical variational algorithm.
The QAOA \cite{farhi2014}, which is a widely studied hybrid algorithm, repeatedly performs a parameterized quantum circuit that updates parameters to decrease the energy. 
Hence, for optimizing all circuit parameters, the QAOA requires multivariate optimization.
This feature is also present in the QAOA with CD driving \cite{chandarana2022}.
In contrast, univariate optimization based on a parameterized quantum circuit \cite{nakanishi2020} optimizes the parameters one at a time.
By repeating the single-parameter optimization, one can deal with a simplified search space in the optimization.
QGO is a type of univariate optimization.
The parameters of the Hamiltonian for QA with CD driving are sequentially determined using the energy gradient, which is obtained from the sensitivity analysis.

Determining the sign of the gradient is a challenge when QGO is implemented on a noisy intermediate-scale quantum (NISQ) \cite{preskill2018} device. 
The gradient can be accurately estimated using two points with a sufficiently small interval in the finite difference approximation.
A smaller interval leads to a smaller difference in the energies of the two points. 
For a NISQ device, identifying a small energy difference is a challenging task because the estimated energies include device and shot noise.
Therefore, as QGO is targeted to run on NISQ devices \cite{kadowaki2021}, a noise-resilience algorithm that can determine the gradient sign is required.

In this paper, we demonstrate a calculation technique for determining the signs of the CD term's parameters for QGO implemented on quantum computers, including NISQ devices.
To ensure the validity of the sensitivity analysis under noise in quantum computing, we adjust the interval to maintain a sufficiently large energy difference in the calculation of the approximated gradient.
By compromising the accuracy of the gradient value, our approach can determine the sign while reducing the number of shots.
We implement QGO with the adjusted interval on a quantum circuit simulator and a quantum computer.
Here, QGO based on the discretization of time evolution is referred to as discretized QGO (d-QGO).
We apply d-QGO to spin glass problems and confirm that it can find the ground state with a small number of measurements in a short annealing time.

The rest of this paper is organized as follows.
Section 2 describes the implementation of d-QGO on a quantum circuit and explains the strategy used for sign determination.
Section 3 presents the results of d-QGO with the improved sign determination implemented on a quantum circuit simulator and a quantum computer. 
Finally, Section 4 summarizes the results and discusses directions for future research.

\section{Methods}

\subsection{Quantum greedy optimization}

QGO employs the same Hamiltonian as that for QA with CD driving \cite{takahashi2017}.
The Hamiltonian of an $n$-qubit system is given by 
\begin{equation}
	\mathcal{H}=A(t)\mathcal{H}^z + B(t)\mathcal{H}^x + \sum_{i=1}^n C_i(t) \mathcal{H}_i^y, \label{eq:QGO_ham}
\end{equation}
where 
\begin{equation}
	\mathcal{H}^{z}=-\sum_{i<j}J_{ij}\sigma_i^z\sigma_j^z,~
	\mathcal{H}^{x}=-\sum_{i=1}^n \sigma_i^x,~
	\mathcal{H}^{y}_i=-\sigma_i^y.
\end{equation}
$\mathcal{H}^x$ and $\mathcal{H}^z$ are the transverse field ($x$-field) Hamiltonian and the Ising (problem) Hamiltonian, and $A(t)$ and $B(t)$ are their time-dependent parameters, respectively.
$\mathcal{H}^y_i$ and $C_i(t)$ are the $y$-field Hamiltonian and its time dependent parameter of the $i$th qubit, respectively.
$\sigma^x$, $\sigma^y$, and $\sigma^z$ are the Pauli $X$, $Y$, and $Z$ matrices, respectively.
The first two terms correspond to the Hamiltonian used in the conventional QA.
The third term is called the CD term, which speeds up the time evolution of the quantum system.

In this paper, we define $A(t), B(t)$, and $C(t)$ as time-dependent functions \cite{prielinger2021,kadowaki2021}:
\begin{equation}
	\begin{split}
		A(t)&=\frac{at}{T}, ~
		B(t)=b\left(1-\frac{t}{T}\right),\\
		C_i(t)&= c_i \sin^2\left(\frac{\pi t}{T}\right),    
	\end{split} \label{eq:schedule_cdd}
\end{equation}
where $T$ is the annealing time and $a, b$, and $c_i$ are parameters that determine the annealing schedule.

\begin{figure}[htbp]
	\includegraphics[keepaspectratio, width=80mm]{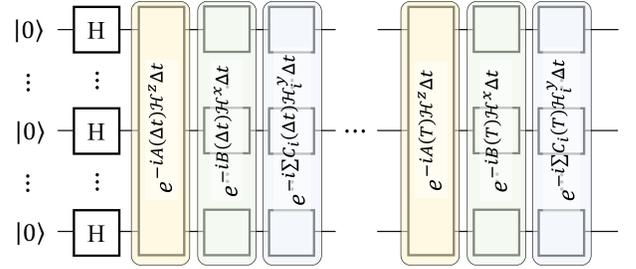}	
	\caption{\label{fig:qc} Schematic representation of a quantum circuit for Hamiltonian \eqref{eq:QGO_ham}. H represents a Hadamard gate. After the Hadamard gates, for the calculation of the time evolution of $\mathcal{H}$ during annealing time $T$, the circuits corresponding to the $\mathcal{H}^x$, $\mathcal{H}^y$, and $\mathcal{H}^z$ terms defined in Eq.~\eqref{eq:time_evolv_CDD_raw} are repeated $M$ times. }
\end{figure}

For the implementation of the Hamiltonian $\mathcal{H}$ on a quantum circuit, we approximate the time evolution operator $\mathcal{U}$ of $\mathcal{H}$ using the Suzuki-Trotter decomposition \cite{suzuki1976}.
Let $\Delta t$ denote the time interval for the time discretization of the Schr\"{o}dinger equation. Then, $\mathcal{U}$ in time $t\in [0, T]$ can be expressed as follows (see Appendix~\ref{appdix:QC} for details):
\begin{equation}
	\begin{split}
		\mathcal{U}(T,0) & = \mathcal{T} \exp \left( - i \int_0^T \mathcal{H}(t) \mathrm{d}t \right) \\
		& = \mathcal{U}(T,T-\Delta t) \mathcal{U}(T-\Delta t, T-2\Delta t) \cdots \mathcal{U}(\Delta t,0),
	\end{split} \label{eq:time_evolv_CDD_raw}
\end{equation}
where 
\begin{equation*}
	\begin{split}
		\mathcal{U}(t+\Delta t, t)&\approx \mathcal{U}_y(t)\mathcal{U}_x(t)\mathcal{U}_z(t), \\
		\mathcal{U}_x(t)&=\exp \left(-i B(t) \mathcal{H}^x \Delta t \right), \\
		\mathcal{U}_y(t)&=\exp \left(-i \sum_{i=1}^n C_i(t) \mathcal{H}_i^y \Delta t   \right), \\
		\mathcal{U}_z(t)&=\exp \left(-i A(t) \mathcal{H}^z \Delta t \right).
	\end{split}
\end{equation*}
$\mathcal{T}$ denotes the time-ordered product and $\Delta t$ is the time step. 
$\Delta t$ is related to the Trotter number $M$, namely $M = T / \Delta t$.
In this paper, we set $\Delta t$ to $0.1$ and thus $M$ increases with $T$.
Figure~\ref{fig:qc} shows a schematic representation of the decomposed Hamiltonian implemented on a quantum circuit.

In QGO, the signs of $c_i$ are determined based on the sensitivity analysis and correspond to the ground state.
Algorithm~\ref{table:QGO} shows the algorithm of QGO.
The parameter $a$ for $\mathcal{H}^x$ is set to 1 for simplicity.
The parameter $b$ for $\mathcal{H}^z$ and the absolute value of $c_i$ for $\mathcal{H}^y_i$ are fixed at constant values, namely $b_\text{opt}^n$ and  $c_\text{opt}^n$, respectively.
In a previous work \cite{kadowaki2021}, it was shown that $b_\text{opt}^n$ and $c_\text{opt}^n$ optimized for a ferromagnetic model with system size $n$ can be used for other optimization problems.
In this paper, we follow the same procedure to determine $b_\text{opt}$ and $c_\text{opt}$.
For calculating the signs of $\mathbf{c}=\{c_1, \cdots, c_n\}$, we utilize the gradient of energy $E(\mathbf{c})$ with respect to $\mathbf{c}$.
The gradient, $\nabla E(\mathbf{c})$, is approximately obtained using the finite difference method:
\begin{equation}
	\begin{split}
		\nabla E(\mathbf{c})&=\left(g_1, \ldots, g_j, \ldots, g_n\right), \\
		g_j & \approx \frac{E(c_1,\ldots,c_j+\Delta c, \ldots, c_n)-E(\mathbf{c})}{\Delta c}, \\
		E(\mathbf{c})&=\braket{\psi(T)|\mathcal{H}^z|\psi(T)},
	\end{split}\label{eq:QGO_diff}
\end{equation}
where $\Delta c$ is the differentiation interval and $\psi(T)$ is the quantum state after the time evolution to annealing time $T$.
The sign of $c_i$ is determined based on the signs of the gradient.
The optimal sign of a given $c_i$ is searched for while the other parameters, $c_j~(j \neq i)$, are fixed.
In this paper, a quantum circuit simulator or a quantum computer is utilized to estimate $E(\mathbf{c})$.

\begin{table}[htbp]
	\begin{algorithm}[H]
		\caption{\label{table:QGO} QGO algorithm \cite{kadowaki2021}}
		\begin{algorithmic}[1]
			\REQUIRE optimized parameters in CD Hamiltonian for $n$-qubit system $c_\text{opt}^n$, differential interval $\Delta c$. 
			\ENSURE solution to target problem
			\STATE $\mathbf{c}\leftarrow\{0,\ldots, 0\}$
			\REPEAT
			\FOR{$j \in \{j|c_j = 0\}$}
			\STATE $g_j \leftarrow (E(c_1,\ldots,c_j+\Delta c, \ldots, c_n) - E(\mathbf{c}))/\Delta c$
			\ENDFOR
			\STATE $i \leftarrow \arg \max_{k|c_k=0} \left| g_k \right|$
			\STATE $c_i \leftarrow -c_\text{opt}^n~\text{sgn}\left(g_i\right)$
			\UNTIL{$c_i \neq 0$ for all $i$}
			\RETURN $ \text{sgn}\left(\mathbf{c}\right)$
		\end{algorithmic}
	\end{algorithm}
\end{table}

\subsection{Parameter settings for sensitivity analysis on NISQ device}

When the value of $\Delta c$ in Eq.~\eqref{eq:QGO_diff} is as small as possible, the gradient $g_j$ can be accurately obtained.
The energy difference (i.e., the numerator of $g_j$) decreases with $\Delta c$.
Regardless of the amount of energy difference, a fault-tolerant quantum computer can calculate the gradient as well as a classical computer.
However, in the current quantum computer, i.e., a NISQ device, it is difficult to accurately identify a small energy difference due to device and shot noise.

\begin{figure}[h]
	\begin{tabular}{c}
		\begin{minipage}[t]{0.9\columnwidth}
			\centering
			\includegraphics[keepaspectratio, scale=0.22]{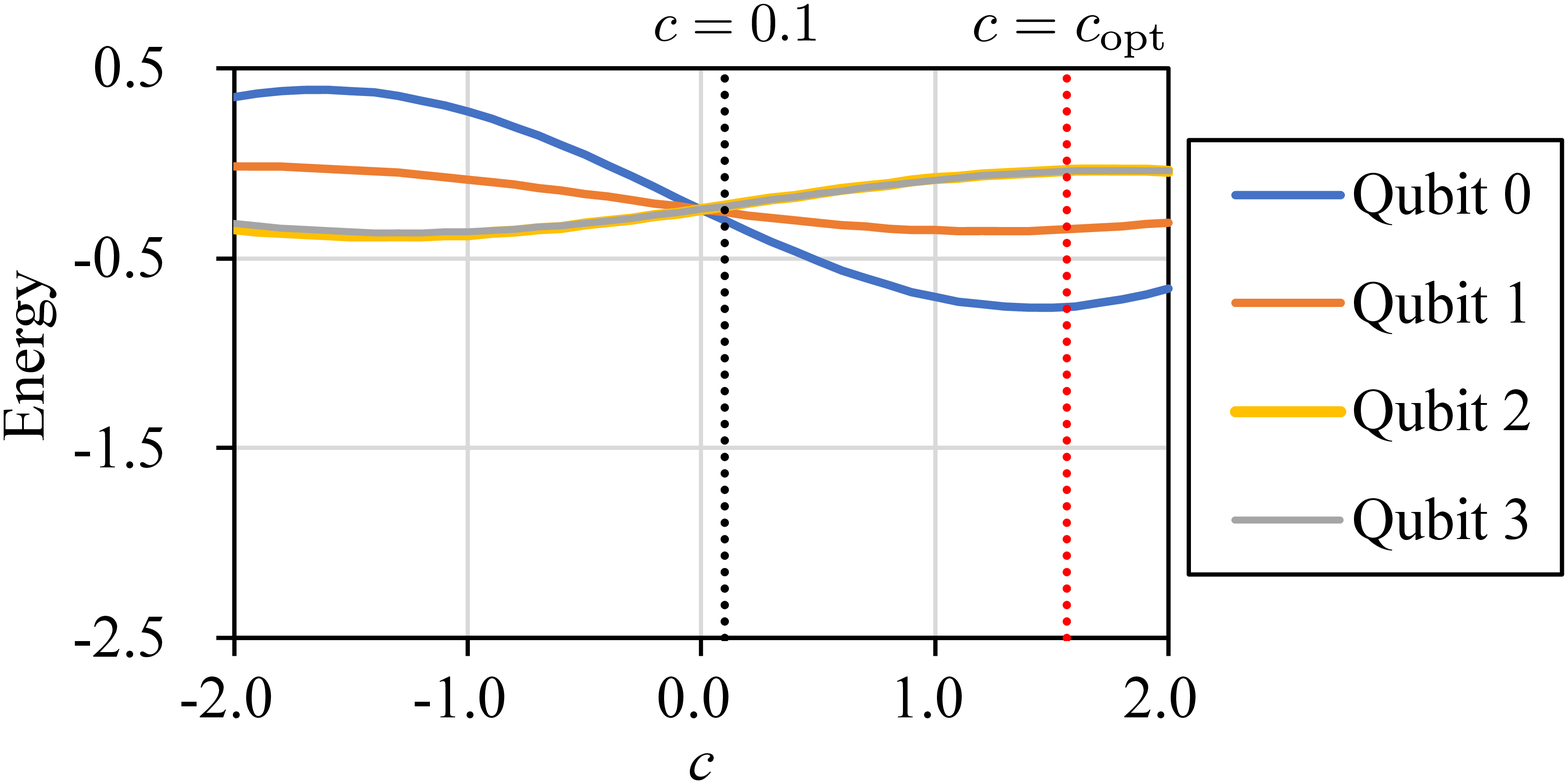}
		\end{minipage} 
		\\
		\\
		\begin{minipage}[b]{0.9\columnwidth}
			\centering
			\includegraphics[keepaspectratio, scale=0.22]{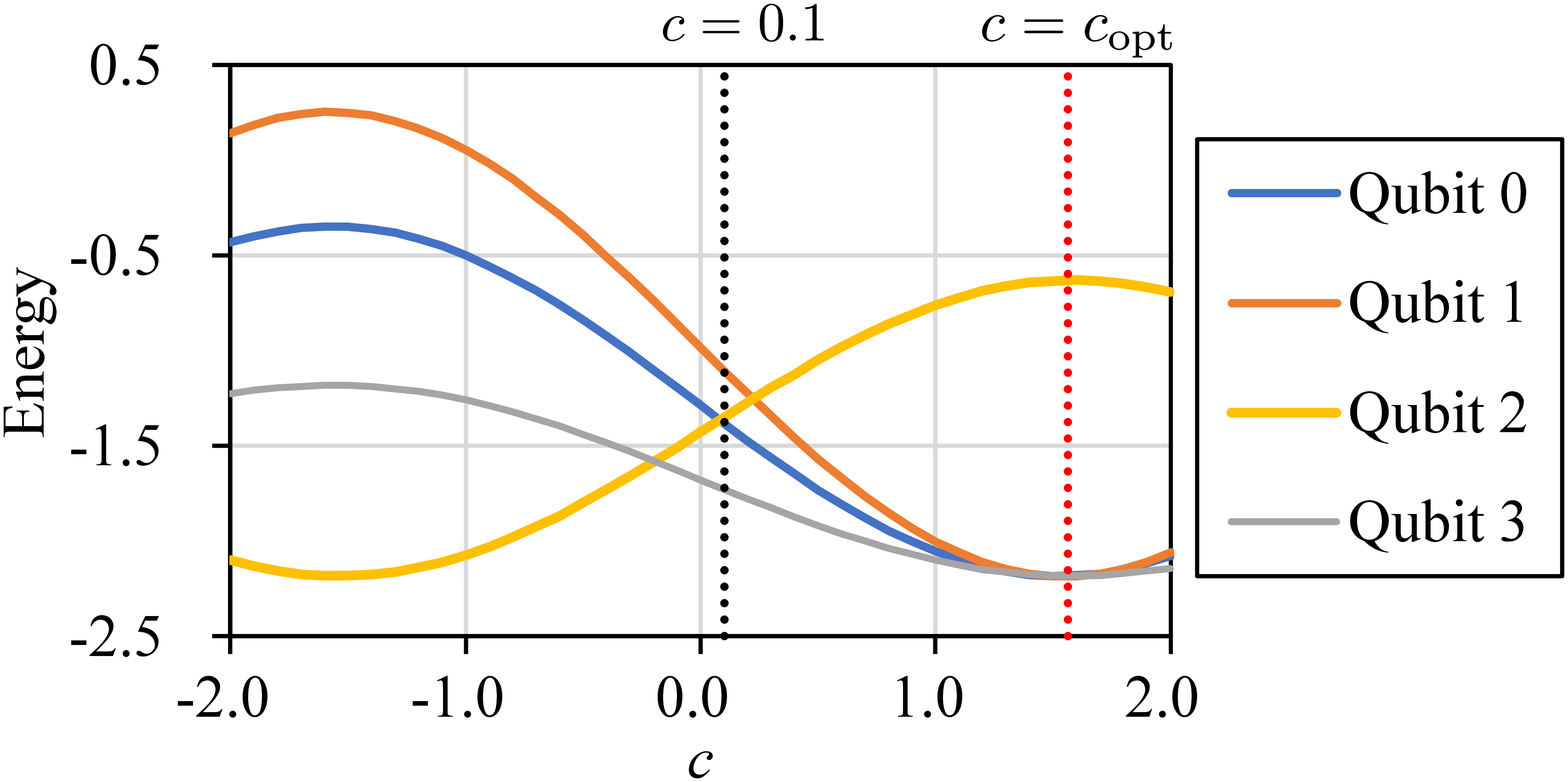}	
		\end{minipage}
	\end{tabular}
	\caption{\label{fig:sim_energy} Example of energy landscapes for $4$-qubit system in first ($\mathbf{c}=\{0,0,0,0\}$, top) and last ($\mathbf{c}=\{-c_\text{opt},-c_\text{opt},+c_\text{opt},-c_\text{opt}\}$, bottom) iterations of QGO. $c_\text{opt}$ for this system is 1.56 (red dashed line). The black dashed line is a vertical line at $c=0.1$. The colored solid lines correspond to energies calculated for $c_i$ in the range of $[-2,2]$ while $c_j$, $j\neq i$, were fixed.}
\end{figure}

To reduce the noise effect to obtain the signs of the gradient, we set $\Delta c=c_\text{opt}$, which provides a larger energy difference than that obtained with a small interval (e.g., $\Delta c=0.1$).
Therefore, when the energy difference remains large enough to avoid energy variance due to noise, the signs can be accurately calculated by compromising the accuracy of the gradient values.
Figure~\ref{fig:sim_energy} shows an example of energy landscapes for a 4-qubit system within $c_i \in [-2,2]$. 
The energy in the first iteration of QGO varies in a narrower range (top of Fig.~\ref{fig:sim_energy}) compared with that in the last iteration (bottom of Fig.~\ref{fig:sim_energy}).
If the gradient is calculated using $\Delta c=0.1$, more measurements are required to retrieve a signal from noise compared with those for the case with $\Delta c = c_\text{opt}$.
Since QGO is a greedy approach, once the gradient sign is misdetected, QGO cannot find the ground state.
When one selects $c_\text{opt}=1.56$ for $\Delta c$, the energy at $c_\text{opt}$ is larger than that for $\Delta c=0.1$.
Therefore, it is expected that a lower or higher energy will be obtained by employing $c_\text{opt}$ for $\Delta c$ instead of a smaller value (e.g., 0.1).

A larger energy difference leads to a fewer measurements.
In quantum computing, due to the probabilistic behavior of the quantum system, one has to run the quantum circuit and measure the quantum state many times.
It is known that the estimation accuracy of the energy depends on the number of measurements (shots).
QGO requires the gradient sign, not the gradient value itself. 
Therefore, if the difference between two energies, $E(c_1)-E(c_2)$, is sufficiently large to identify the sign under noisy conditions, the required number of shots for estimating the sign can be reduced.

\section{Results} \label{sec:results}
\subsection{Differentiation interval effect in QGO}

We applied d-QGO to find the ground state in the Sherrington-Kirkpatrick (SK) model with the local field defined in Appendix~\ref{appdix:SKmodel}. 
Figure~\ref{fig:sim_sp_T} shows success probabilities (SPs) of d-QGO when the system size is varied from $n$ = 4 to 16 for annealing times ($T=1.0$ and $10.0$).
The SP for QGO is defined as the probability of finding the ground state among 100 instances.
In Fig.~\ref{fig:sim_sp_T}, we also show SP of AQC based on the discretization of time evolution (d-AQC), which corresponds to d-QGO without the CD term, and vanilla QGO (v-QGO), which is QGO implemented on the quantum simulator QuTiP \cite{johansson2012}.
Note that since d-AQC does not have the CD term, it requires a longer annealing time than that for QGO to find the ground state.
Therefore, we employed annealing time $T=10.0$ for d-AQC and $T=1.0$ for v-QGO.

The proposed algorithm, d-QGO, had a higher SP and a shorter annealing time than those of d-AQC independent of the system size.
The SP of d-QGO with $T=1$ is comparable to that of v-QGO.
As the annealing time increased, the SP of d-QGO decreased to half that of d-AQC.
This result is consistent with the result in a previous work \cite{kadowaki2021}, in which QGO found the ground state in a shorter annealing time compared with that for d-AQC.

\begin{figure}[h]
	\includegraphics[keepaspectratio, scale=0.7]{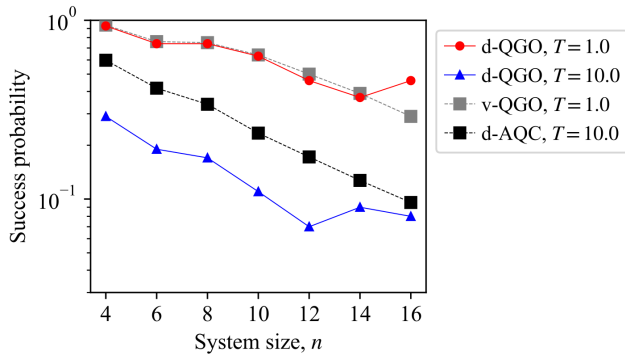}
	\caption{\label{fig:sim_sp_T} System size dependence of SP of d-QGO (solid lines) with annealing time for $T=1$ and $10$  for system sizes of $n=4$ to $16$. The SPs of v-QGO (dashed grey line) and d-AQC (dashed black line) are also plotted.  d-QGO and d-AQC were implemented on the quantum gate simulator Qulacs \cite{suzuki2021} and v-QGO was implemented on the quantum simulator QuTiP \cite{johansson2012}. }
\end{figure}

\begin{figure}[h]
	\includegraphics[keepaspectratio, scale=0.7]{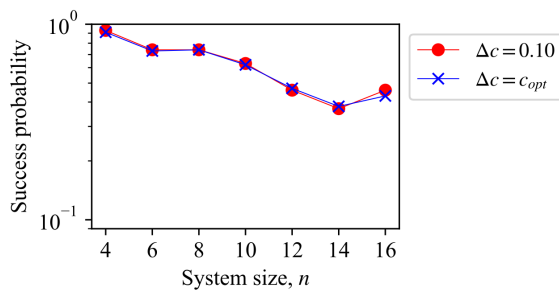}
	\caption{\label{fig:sim_sp_dc} SP comparison for differential intervals $\Delta c=0.1$ and $c_\text{opt}$ for annealing time of $T=1.0$. }
\end{figure}

Changing the differentiation interval $\Delta c$ in the short annealing time $T=1$ does not affect SP, as shown in Fig.~\ref{fig:sim_sp_dc}.
For the system sizes considered here, $c_\text{opt}$ is always larger than 0.1.
The SP is not significantly affected by an increase in $\Delta c$.
This result indicates that when $\Delta c$ takes a sufficiently large value (e.g., $c_\text{opt}$), the sign of the energy gradient, $\text{sgn}(g_i)$, can be accurately estimated even if d-QGO is implemented on a NISQ device. 

\subsection{QGO on NISQ device}

We implemented d-QGO on the quantum computer ibmq\_lima, which is based on the IBM Quantum Falcon Processor \cite{IBMQ}.
We prepared 50 instances for the 2-qubit system.
In this subsection, the annealing time, $T$, was set to $1.0$.
The depth of the circuit for the SK model Hamiltonian in the 2-qubit system was 61. After circuit optimization, the depth was decreased to 16.
Note that the optimized circuit depth of the 2-qubit system for any annealing time (any Trotter number) was decreased, resulting in the same depth as that for $T=1.0$.
The optimized quantum circuit was run on the quantum computer.

\begin{table}[b]
	\caption{
		\label{tab:IBMQ_SP} SP of d-QGO on quantum computer. d-QGO with various differentiation intervals ($\Delta c $) and numbers of shots for estimating energy was run on ibmq\_lima. d-AQC on the quantum computer (ibmq\_lima (d-AQC)) and d-QGO on the quantum circuit simulator (ibmq\_qasm\_simulator) were run for a comparison of SP for d-QGO. Note that in the 2-qubit system, $c_\text{opt}=1.523$.
	}
	\begin{ruledtabular}
		\begin{tabular}{ccrc}
			Device (algo.) & $\Delta c$ & Shots & SP  \\
			\colrule
			ibmq\_lima (d-AQC) &          & 10000 & 30.0 \% \\ 
			ibmq\_lima & 0.10 &   2000 & 76.0 \% \\
			ibmq\_lima & 0.10 & 10000 & 96.0 \% \\
			ibmq\_lima & 1.523 & 2000   & \textbf{98.0} \% \\
			ibmq\_lima & 1.523 & 10000 & \textbf{98.0} \% \\
			ibmq\_qasm\_simulator & 1.523 & 10000 & \textbf{98.0} \% \\
		\end{tabular}
	\end{ruledtabular}
\end{table}

\begin{figure}[htbp]
	\includegraphics[keepaspectratio, scale=0.7]{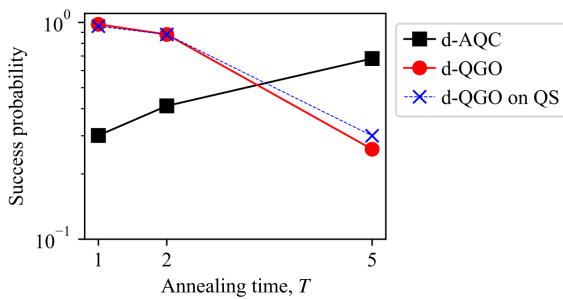}
	\caption{\label{fig:IBMQ_SP_time} Annealing time dependence of SP on quantum computer, d-AQC (10000shots) and d-QGO ($\Delta c=c_\text{opt}$, 2000 shots), and quantum simulator, d-QGO on QS ($\Delta c=0.1$, 10000 shots). Note that the circuit depth depends on annealing time; however, in this case, the optimized circuit depth is always 16 (including two CX gates) independent of annealing time.}
\end{figure}

The number of shots for estimating the energy in d-QGO on a quantum computer can be reduced by enlarging the differential interval without decreasing the SP.
Table~\ref{tab:IBMQ_SP} shows SPs obtained for differentiation intervals ($\Delta c = 0.1$ and $c_\text{opt}=1.523$) and numbers of shots (2000 and 10000).
The SP of d-AQC with a short annealing time is 30\%.
For d-QGO with $\Delta c=0.1$, increasing the number of shots from 2000 to 10000 leads to a higher SP (from 76\% to 96\%). 
However, the SP of d-QGO with $\Delta c=c_\text{opt}$ is constant at 98\%, independent of the number of shots. This value is identical to that obtained with the simulator.
Figure~\ref{fig:IBMQ_SP_time} shows the SPs of d-AQC and d-QGO implemented on the quantum computer and d-QGO implemented on the quantum simulator.
The SP of d-QGO, for both the computer and simulator, decreases with increasing annealing time, whereas that of d-AQC increases.
This result is consistent with the results of the quantum circuit simulation. 

\section{Discussion}

In this paper, we presented QGO based on a discretization of the time evolution implemented within a quantum circuit framework (d-QGO). 
The QGO determines the sign of the counterdiabatic (CD) term based on a sensitivity analysis of energy.
Estimating the gradient sign correctly is crucial for finding the ground state in QGO, as the system follows a temporal ground state guided by the CD driving.

The gradient sign estimated by quantum computing, however, includes calculation errors due to device and shot noise. 
To ensure the correct estimation of the sign, we employed a larger differential interval in the sensitivity analysis, consequently increasing the energy difference. 
This enlarged energy difference not only improves the probability of finding the ground state but also reduces the number of required measurements (shots).

We investigated the performance of d-QGO in terms of finding the ground state on a quantum circuit simulator and a noisy quantum computer. 
Our results using a quantum circuit simulator (Fig.~\ref{fig:sim_sp_T}) indicate that both d-QGO and vanilla QGO (v-QGO), proposed in a previous work \cite{kadowaki2021}, outperform AQC based on the discretization of time evolution (d-AQC) in a short annealing time on a 16-qubit system. 
Moreover, we demonstrated that d-QGO implemented on the quantum computer has a higher SP with a shorter annealing time than that of d-AQC (Table~\ref{tab:IBMQ_SP}); the SP of d-AQC increases with annealing time (Fig.~\ref{fig:IBMQ_SP_time}). 
These results suggest that QGO could overcome the drawback of d-AQC with respect to annealing time when implemented on a quantum computer.

The results of the quantum circuit simulation and the quantum computer experiment show the potential of d-QGO to reduce the number of required shots. 
Using the quantum circuit simulator, we confirmed that SP is not affected by an increase in the differential interval (Fig.~\ref{fig:sim_sp_dc}).
For a 2-qubit system, we demonstrated that d-QGO produced the same SP on the quantum computer as that on the quantum circuit simulator, and that it reduced the number of required shots (Table~\ref{tab:IBMQ_SP}).
The circuit depth for d-QGO increases with system size, which, in general, requires a large number of shots to maintain precise estimation in the presence of noise.
Further studies are needed to confirm the advantage of d-QGO in reducing the number of shots required for many-qubit systems and other optimization problems.

The QAOA \cite{farhi2014} and QGO \cite{kadowaki2021} are quantum variational algorithms for optimization.
The QAOA and its extensions to CD driving \cite{chandarana2022,wurtz2022} require multivariate optimization of circuit parameters, whereas QGO and d-QGO iteratively perform univariate optimization of circuit parameters.
As univariate optimization can simplify the search space compared with that for multivariate optimization, it is expected that the simplification contributes to the high SP of d-QGO.
However, the performance of the QAOA and d-QGO depends on the circuit depth.
In d-QGO, the circuit depth is proportional to the annealing time and the Trotter number used in the decomposition of the Hamiltonian (Eq.~\eqref{eq:time_evolv_CDD_raw}), in which the CD term generates an additional circuit.
Therefore, the two algorithms should be compared in terms of circuit depth and SP in future work.

\appendix

\section{\label{appdix:QC} Quantum circuit for counterdiabatic driving in quantum annealing} 

The time evolution of a quantum system is expressed by the Schr\"{o}dinger equation:
\begin{equation}
	i\frac{\partial}{\partial t}\ket{\psi(t)} = \mathcal{H} \ket{\psi(t)}, \label{eq:sh}
\end{equation}
where $\ket{\psi(t)}$ and $\mathcal{H}$ are respectively the quantum state vector and the Hamiltonian of the target quantum dynamics.
The Hamiltonian of an $n$-qubit system for counterdiabatic driving algorithms is given by Eq.~\eqref{eq:QGO_ham}, which is repeated below.
\begin{equation}
	\mathcal{H}=A(t)\mathcal{H}^z + B(t)\mathcal{H}^x + \sum_{i=1}^n C_i(t) \mathcal{H}_i^y,
\end{equation}
where 
\begin{equation*}
	\mathcal{H}^{z}=-\sum_{i<j}J_{ij}\sigma_i^z\sigma_j^z,~
	\mathcal{H}^{x}=-\sum_{i=1}^n \sigma_i^x,~
	\mathcal{H}^{y}_i=-\sigma_i^y.
\end{equation*}

Let us define the time evolution operator $\mathcal{U}$ of the Schr\"{o}dinger equation \eqref{eq:sh} for expressing a quantum state in the annealing process.
Then, the time evolution from a given initial state $\ket{\psi(0)}$ to a final state $\ket{\psi(T)}$ is expressed by 
\begin{equation}
	\ket{\psi(T)} =\mathcal{U}(T,0)\ket{\psi(0)},
\end{equation}
where $T$ is the annealing time.
For implementing $\mathcal{U}$ for a circuit gate in a quantum computer, one generally approximates $\mathcal{U}$ by discretizing time and then applying the Suzuki-Trotter decomposition. The discretized $\mathcal{U}$ is obtained as shown in Eq.~\eqref{eq:time_evolv_CDD_raw}, which is repeated below.
\begin{equation}
	\begin{split}
		\mathcal{U}(T,0) & = \mathcal{T} \exp \left( - i \int_0^T \mathcal{H}(t) \mathrm{d}t \right) \\
		& = \mathcal{U}(T,T-\Delta t) \mathcal{U}(T-\Delta t, T-2\Delta t) \cdots \mathcal{U}(\Delta t,0),
	\end{split} \label{eq:time_evolv_CDD}
\end{equation}
where 
\begin{equation*}
	\begin{split}
		\mathcal{U}(t+\Delta t, t)&\approx \mathcal{U}_y(t)\mathcal{U}_x(t)\mathcal{U}_z(t), \\
		\mathcal{U}_x(t)&=\exp \left(-i B(t) \mathcal{H}^x \Delta t \right), \\
		\mathcal{U}_y(t)&=\exp \left(-i \sum_{i=1}^n C_i(t) \mathcal{H}_i^y \Delta t   \right), \\
		\mathcal{U}_z(t)&=\exp \left(-i A(t) \mathcal{H}^z \Delta t \right).
	\end{split}
\end{equation*}
$\mathcal{T}$ denotes the time-ordered product and $\Delta t$ is the time step.

The time evolution operator, $\mathcal{U}$, is expressed by quantum gates.
In $\mathcal{U}_x(t)$, $\mathcal{U}_y(t)$, and $\mathcal{U}_z(t)$, the circuits acting on $i$ and $j$ qubits are expressed by
\begin{align} 
	e^{-i \beta \sigma^x_i }&=RX_i(2\beta), \\
	e^{-i \alpha_i \sigma^y_i}&=RY_i(2\alpha_i), \\
	e^{-i \gamma \sigma^z_i \sigma^z_j}&= {CX}_{ij} RZ_j(2\gamma) {CX}_{ij},
\end{align}
where $RX$, $RY$, and $RZ$ are single qubit rotation gates about the X, Y, and Z axis, respectively.
${CX}_{ij}$ is called the controlled X gate, which operates in the same way as the CNOT gate in classical computing.
The parameters $\alpha, \beta$, and $\gamma$ are the corresponding rotation angles, which are defined by
\begin{align} 
	\alpha_i &= C_i(t) \Delta t, \\
	\beta &= B(t) \Delta t, \\
	\gamma&=A(t)J_{ij} \Delta t.
\end{align}
The quantum circuit for $\mathcal{U}$ can be generated by repeating the sets of $\mathcal{U}_x$, $\mathcal{U}_y$, and $\mathcal{U}_z$ $M$ times, as shown in Fig.~\ref{fig:qc}. 

\section{\label{appdix:SKmodel} Sherrington-Kirkpatrick model with local field}
We applied quantum optimization algorithms to the search of the ground state in the Sherrington-Kirkpatrick model with a local field Hamiltonian.
This model is a typical spin glass problem. Its Hamiltonian for an $n$-qubit system is defined as
\begin{equation}
	\mathcal{H}^{z}_J= - \sum_{i<j,i=0}^n\sum_{j=0}^n J_{ij} \sigma_i^z\sigma_j^z, \label{eq:SK_Hz_Jij}
\end{equation}
where $J_{ij}$ is determined based on a probability density function $P$, namely 
\begin{equation}
	P(J_{ij}) = \frac{1}{\sqrt{2\pi \sigma^2}} \exp\left(-\frac{J_{ij}^2}{2\sigma^2}\right), ~\sigma^2=\frac{1}{n}. \label{eq:pdf}
\end{equation}
In section \ref{sec:results}, we adopt the Hamiltonian \eqref{eq:SK_Hz_Jij}, to which the local field Hamiltonian is added, namely
\begin{equation}
	\mathcal{H}^{z} = \mathcal{H}^{z}_J - \sum_{i}^n h_{i} \sigma_i^z, \label{eq:SK_Hz_hi}
\end{equation}
where $h_i$ is determined based on the probability density function given in Eq.~\eqref{eq:pdf}.
This Hamiltonian has a unique ground state.
\newline

\begin{acknowledgments}
	The authors acknowledge Hirotaka Irie for his fruitful discussion.
	We acknowledge the use of IBM Quantum services for this work. The views expressed are those of the authors and do not reflect the official policy or position of IBM or the IBM Quantum team.
	
\end{acknowledgments}

\bibliography{DQGO.bib}

\end{document}